\documentclass[prb,aps,twocolumn]{revtex4}
\usepackage{amsmath}
\usepackage{amssymb}
\usepackage {graphicx}
\usepackage{color}

\newcommand{\beq}{\begin{equation}}
\newcommand{\eeq}{\end{equation}}
\newcommand{\bea}{\begin{eqnarray}}
\newcommand{\eea}{\end{eqnarray}}
\newcommand{\be}{\begin{equation}}
\newcommand{\ee}{\end{equation}}

\newcommand{\e}{\varepsilon}
\newcommand{\bk}{{\vec k}}

\newcommand{\br}{{\vec r}}
\newcommand{\bj}{{\vec j}}

\newcommand{\vj}{{\vec{j}}}

\begin{document}

\title{Spontaneous currents in a superconductor with $s+$i$s$ symmetry }
\author{Saurabh Maiti$^{1,2}$, Manfred Sigrist$^3$, Andrey Chubukov$^4$}
\affiliation{$^1$Department of Physics, University of Florida, P. O. Box 118440, Gainesville, Florida 32611-8440, USA}
\affiliation{$^2$National High Magnetic Field Laboratory, Tallahassee, FL 32310, USA}
\affiliation{$^3$Institut fur Theoretische Physik, ETH Zurich, 8093 Zurich, Switzerland}
\affiliation{$^4$William I. Fine Theoretical Physics Institute, and School of Physics and Astronomy,
University of Minnesota, Minneapolis, MN 55455, USA}
\date{\today}
\begin{abstract}
We analyze $s+is$ state proposed as a candidate superconducting state for strongly hole-doped
Ba$_{1-x}$K$_x$Fe$_2$As$_2$. Such a state breaks time-reversal symmetry (TRS) but does not break any other discrete symmetry. We address the issue whether  TRS breaking alone can generate spontaneous currents near impurity sites, which could be detected in, e.g., $\mu$SR experiments.   We argue that there are no spontaneous currents if only TRS is  broken. However, supercurrents do emerge if the system is put under external strain and $C_4$ lattice rotation symmetry is externally broken.
\end{abstract}
\maketitle
{\it Introduction}~~~~The search for a truly unconventional superconductivity which breaks time-reversal symmetry (TRS) in addition to  $U(1)$ overall phase symmetry of a superconducting order parameter continues to attract
a lot of attention in the physics community.~\cite{Sigrist,Volovikold,Sachdev} Systems which break TRS exhibit a wealth of fascinating properties and are highly sought after for applications~\cite{Kane,Qi,Cheng,Kopnin,yakovenko,laughlin, Volovik, Senthil, Horovitz,Sato,Mao}. The TRS breaking pairing states have been proposed in the two-dimensional ${}^3$He (Ref. \onlinecite{Volovikold}) and for the fractional quantum Hall effect at $5/2$ filling.\cite{Read,ivanov} In solid-state realizations, TRS-breaking (TRSB) $p_x+ip_y$  superconductivity has been proposed for Sr$_2$RuO$_4$  (Ref. ~\onlinecite{Mackenzie})  and found to be in agreement with the measurements of the polar Kerr effect, specifically designed to measure TRS breaking.~\cite{kapit} There are no experimental realizations yet of TRSB spin-singlet superconductivity, although $d_{x^2-y^2} + i d_{xy}$ state has been proposed theoretically for hexagonal systems near van-Hove doping, including doped graphene,\cite{rahul,ronny_1,gonzales} SrPtAs ~\cite{ronny_2} and, possibly, cobaltates.~\cite{ronny_3}

The search for TRS breaking superconductivity intensified with the discovery of Fe-based superconductors (FeSCs). These systems have multiple Fermi surfaces, and intra-pocket and inter-pocket interactions vary along each Fermi surface as the consequence of different orbital compositions of low-energy excitations.~\cite{reviews} This variation opens a possibility that the pairing interaction is attractive in more than one channel. In FeSCs, the two leading candidates are $s^{+-}$ and $d_{x^2-y^2}$ channels. The majority of researchers believe that in moderately doped FeSCs
$s^{+-}$ superconductivity wins, but some experiments on heavily hole-doped system Ba$_{1-x}$K$_x$Fe$_2$As$_2$ with $x =1$ were interpreted in favor of $d_{x^2-y^2}$  superconductivity~\cite{d-wave}. This interpretation is not universally accepted~\cite{shin}, but if  it is correct, then one can expect that there will be a mixed state at $x \leq 1$, in which both $s$ and $d$ components are present. According to calculations~\cite{s+id} relative phase between $s$ and $d$-components is $\pm \pi/2$ (an $s+id$ state). In such a state time-reversal(TR) and C$_4$ lattice rotational symmetry are simultaneously  broken, together with $U(1)$ phase symmetry, and this gives rise to a number of non-trivial properties, including circulating supercurrents near a nonmagnetic impurity.\cite{wu}

In a separate line of research, several groups argued that, even if the gap symmetry in Ba$_{1-x}$K$_x$Fe$_2$As$_2$ remains $s$-wave for all $x$, the multi-pocket nature of FeSCs still allows for non-trivial superconductivity at intermediate $x \leq 1$, which break TRS. The argument is that the $s^{+-}$ order parameter (the gap) at $x=1$ (KFe$_2$As$_2$) is qualitatively different from the one at optimal doping $(x \approx 0.4$), where
both hole and electron pockets are present. At optimal doping the most natural $s^{+-}$ state is the one in which the gaps on all hole pockets have the same sign, opposite to that on electron pockets. At $x =1$  (i.e., in KFe$_2$As$_2$) only hole pockets are present~\cite{boris} and, if the gap remains  $s$-wave, it must change sign between the two inner hole  pockets.\cite{maiti_11} It has been argued\cite{maiti_13} that  the transformation  of one $s-$wave structure into the other is not continuous at low enough $T$ and involves an intermediate state in which the phases of the gaps on the two inner hole pockets differ by $0<\alpha<\pi$ (see Fig. \ref{fig:FS}). Such a state breaks TRS, despite that it has pure $s$-wave symmetry, and was termed $s+is$.

The issue we discuss in this letter is how to detect the $s+is$ state experimentally. One option is to  detect low-energy Leggett-type modes in the $s+is$ state~\cite{maiti_13,lara} by, e.g., Raman measurements. But it would be much more desirable to have experimental probes which would directly detect TRS breaking. In this respect, $s+is$ state presents a challenge. The TRS-broken states studied before break other discrete symmetries in addition TR, e.g., $s + i d_{x^2-y^2}$ state breaks C$_4$ symmetry and $d_{x^2-y^2} + i d_{xy}$  breaks mirror symmetries. The $s+is$ state only breaks TRS but keeps lattice symmetries intact.  We show that in this situation there are no circulating supercurrents near a non-magnetic impurity.  We show, however, that supercurrents do develop if a system with $s+is$ gap is put under external strain that breaks the C$_4$ lattice rotational symmetry. Our results are  in agreement with recent zero-field muon-spin relaxation study of superconducting Ba$_{1-x}$K$_x$Fe$_2$As$_2$ for $0.5 <x<0.9$ (Ref. [\onlinecite{sonier}]). The measurements on polycrystalline samples, which do not exhibit the mesoscopic phase separation, showed no evidence of spontaneous internal magnetic fields at temperatures down to 0.02 K.  We propose to preform the same $\mu$SR measurement under external strain.

{\it $s+id$ superconductor~~} To set the stage for our analysis of $s+is$ state it is instructive to consider first $s+id$ state for which numerical calculations~\cite{wu} have shown the presence of spontaneous currents around an inhomogeneity. This will help us understand  the distinction between $s+id$ and $s+is$ states and set up necessary conditions for the existence of currents in an $s+is$ superconductor. The free energy for a candidate $s+id$ system can be written as the combination of the homogeneous and spatially varying parts: $\mathcal{F}=\mathcal{F}_h+\mathcal{F}_s$, where
\bea\label{eq:sidfree}
&&\mathcal{F}_{h}= \alpha_s
|\Delta_s|^2 + \alpha_d |\Delta_d|^2 + \beta_1|\Delta_s|^4 + \beta_2|\Delta_d|^4\nonumber\\
&& ~~~+ \beta_3|\Delta_s|^2|\Delta_d|^2+\beta_4\left( \Delta_s^*\Delta_s^*\Delta_d\Delta_d~+~\text{c.c.} \right)\nonumber\\
&&\mathcal{F}_{s}= \gamma_s |\vec{D}\Delta_s|^2 + \gamma_d|\vec{D}
\Delta_d|^2 \nonumber\\
&&~~~+\gamma_{sd} \left[(\vec{D}_x \Delta_s)^*\vec{D}_x \Delta_d- (\vec{D}_y \Delta_s)^*\vec{D}_y\Delta_d~+~\text{c.c.}\right].
\eea
Here $\Delta_d$ is the magnitude of $d_{x^2-y^2}$ gap ($=\Delta_d \cos 2 \theta$) and the integration over $\theta$ is already carried out. The two order parameters are  $U(1)$ fields $\Delta_s = \Delta e^{i \phi_s}$ and $\Delta_d = \Delta e^{i\phi_d}$. In the  spatially varying part $\vec{D}\equiv -i\vec{\partial} -\frac{2e}{c}\vec{A}$ and all derivatives act of the center of mass co-ordinate of the Cooper pair.~\cite{comm}
We assume that the parameters $\alpha_i$ and $\beta_i$ of the homogeneous part $\mathcal{F}_{h}$ are such that the ground state is a TRSB $s+id$ superconductor with $\phi_d-\phi_s =\pm \pi/2$.

The term with the prefactor $\gamma_{sd}$ depends on the relative phase $\phi_s - \phi_d$ of the two order parameters, but not on the cumulative phase $\phi_s + \phi_d$. This term is  consistent with the symmetry of $s+id$ state  as it remains invariant if one  changes the relative phase $\phi_d - \phi_s$ by $\pi$  and simultaneously rotates the reference frame by $90^o$. Both symmetry operations change $\Delta_d \to -\Delta_d$, and $\mathcal{F}$
is invariant under the product of these two operations.

We now show that the $\gamma_{sd}$ term in the free energy (\ref{eq:sidfree}) gives rise to circulating currents around inhomogeneities. For this, we introduce an isotropic impurity  at $\bf{r}=0$, obtain coordinate-dependent $\Delta_s({\bf r})$ and   $\Delta_d({\bf r})$,  and compute the  current density ${\bj} = -\frac{\partial \mathcal{F}_s}{\partial {\vec{A}}}\left.\right|_{\vec{A}=0}$. We follow Ref. ~\onlinecite{wu} and assume that an impurity introduces  coordinate dependencies of the prefactors $\alpha_i$ in (\ref{eq:sidfree}) via $\alpha_i \rightarrow \alpha_i +\alpha_{\text{imp,i}}(\vec{r})$ ($i =s,d$), where $\alpha_{\text{imp,i}}(\vec{r})$ is a decreasing function of $r$. We take the same $\alpha_{\text{imp,i}}=\alpha_{0,i} e^{-(r/r_{0,i})^2}$ as in Ref. \onlinecite{wu}, but the results would be qualitatively similar for any isotropic impurity potential.
To simplify the presentation, we assume $\alpha_s = \alpha_d$ and drop the index $i$.

\begin{figure}
$\begin{array}{c}
\includegraphics[width=1\columnwidth]{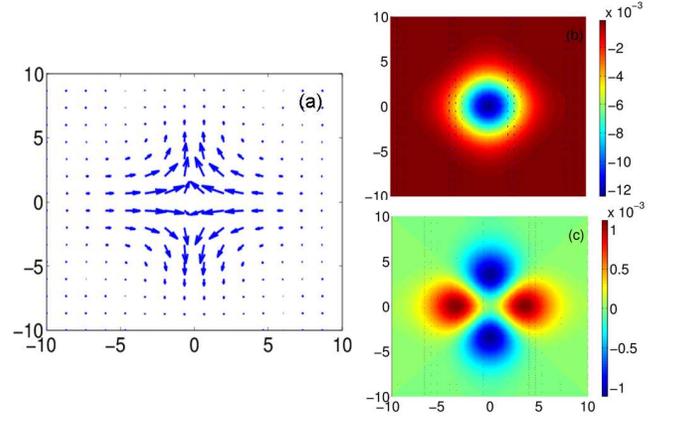}
\end{array}$
\caption{(a)The profile of supercurrent (blue arrows) in real space in the $s+id$ superconductor around an isotropic impurity located at the origin. The current satisfies the continuity equation and creates a magnetic quadrupole moment centered on the impurity site. We set $\tilde\gamma=1$, sample size $L=10r_0$, $\delta=0.5/r_0$. (b) and (c) are the amplitude and phase fluctuations, $\gamma_s m_s=\gamma_d m_d$ and $\sqrt{\gamma_s\gamma_d}\phi_d=-\sqrt{\gamma_s\gamma_d}\phi_d$ , respectively. Note the $d-$wave symmetry of phase fluctuations.}
\label{fig:curra}
\end{figure}

\begin{figure}
$\begin{array}{c}
\includegraphics[width=0.9\columnwidth]{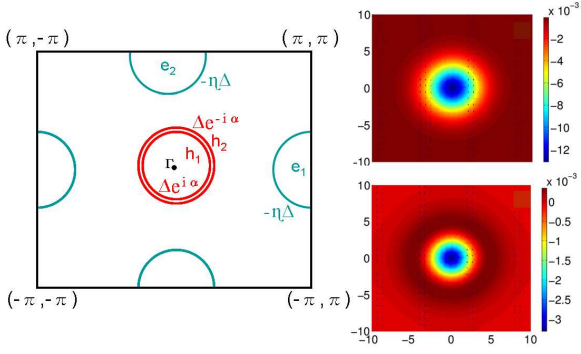}
\end{array}$
\caption{Left: The Fermi surface of the microscopic model considered in the text. It has two hole pockets around the $\Gamma$ point and two electronic pockets at $(\pi,0)$ and $(0,\pi)$  points of the Brillioun zone. We tune the parameters of this model to the TRSB state and expand the gaps in spatial fluctuations around the impurity site in  TRSB $s+is$ state to obtain the currents. Right:  The spatial variation of the amplitude ($m_1$, top) and the phase ($\phi_1$, bottom) of the gap on one of hole pockets without external strain. Observe that there is no phase modulation unlike in $s+id$ state.}
\label{fig:FS}
\end{figure}

We treat $\alpha_{\text{imp}}(\vec{r})$ as a small perturbation and linearize the response around the TRSB state by setting  $\phi_s =0$ and $\phi_d = \pi/2$ away from impurity and expanding $\Delta_s$ and $\Delta_d$ as
$\Delta_s=\Delta(1 + m_s + i\phi_s)$  and $\Delta_d=i\Delta(1 + m_d + i\phi_d)$,
where $m_s, m_d, \phi_s, \phi_d = \mathcal{O}(\alpha_0)$ are all functions of $\vec{r}$. Minimizing $\mathcal{F}$ with respect to variations of $\Delta_s$ and $\Delta_d$ we obtain~\cite{comm}
\bea\label{eq:solutions1}
\gamma_d m_d = \gamma_s m_s &=& -\frac12\left(t_+ + t_-\right), \nonumber\\
\sqrt{\gamma_s\gamma_d} \phi_d =-\sqrt{\gamma_s\gamma_d} \phi_s&=& \frac12 \left( t_+ - t_- \right),
\eea
where
\beq
t_{\pm} \equiv t_{\pm}  (r)
 = \int\frac{d^2 k}{(2\pi)^2} \frac{e^{i{\bk}\cdot {\br}}\tilde\alpha_{\text{imp}}}{k_x^2(1\pm\tilde\gamma) + k_y^2(1\mp\tilde\gamma) + \delta^2}.
 \label{n_1}
\eeq
In Eq. (\ref{n_1})  $\tilde\alpha_{\text{imp}}$
is the Fourier transform of $\alpha_{\text{imp}}$, $\tilde\gamma \equiv \gamma_{sd}/\sqrt{\gamma_s\gamma_d}$, and $\delta$ is the mass scale set by $\beta_i$ in (\ref{eq:sidfree}).
The current density is expressed as
\beq\label{eq:cur}
\bj=-j_0\left[\gamma_s\vec{\partial}\phi_s + \gamma_d\vec{\partial}\phi_d + \gamma_{sd} \vec{\partial}_d(m_d-m_s)\right],
\eeq
where $\vec{\partial}_d=\hat{x}\partial_x  - \hat{y}\partial_y $. Substituting the results for $m_s, m_d, \phi_s, \phi_d$ from (\ref{eq:solutions1}) we obtain
\beq\label{eq:currents2}
\bj=\frac{j_0}{2}\frac{\gamma_s-\gamma_d}{\sqrt{\gamma_s\gamma_d}}\left[\vec{\partial}(t_+-t_-) + \tilde\gamma\vec{\partial}_d(t_++t_-)\right].
\eeq
One can easily verify that $\bj$ satisfies the continuity equation $\vec\nabla\cdot\bj =0$. The difference  $t_+-t_-$ contains  ${\tilde \gamma} \propto \gamma_{sd}$ as the overall factor (see Eq. (\ref{n_1}), hence both terms in the square brackets in (\ref{eq:currents2}) scale with
$\gamma_{sd}$, and $|{\vec j}| \propto \gamma_{sd}$. We see that the current  is entirely due to the mixed $\gamma_{sd}$ term in (\ref{eq:sidfree}), which, we remind, remains invariant under the product of $C_4$ rotation and TR. The presence of two symmetry transformations, each of which changes $\Delta_d$ to $-\Delta_d$, is therefore crucial to obtain a non-zero circulating current around an impurity.  The current ${\bf j} ({\bf r})$ is plotted in Fig. \ref{fig:curra} and agrees well with the numerical results.~\cite{wu}

\begin{figure*}\label{fig:currb}
$\begin{array}{cc}
\includegraphics[width=0.75\columnwidth]{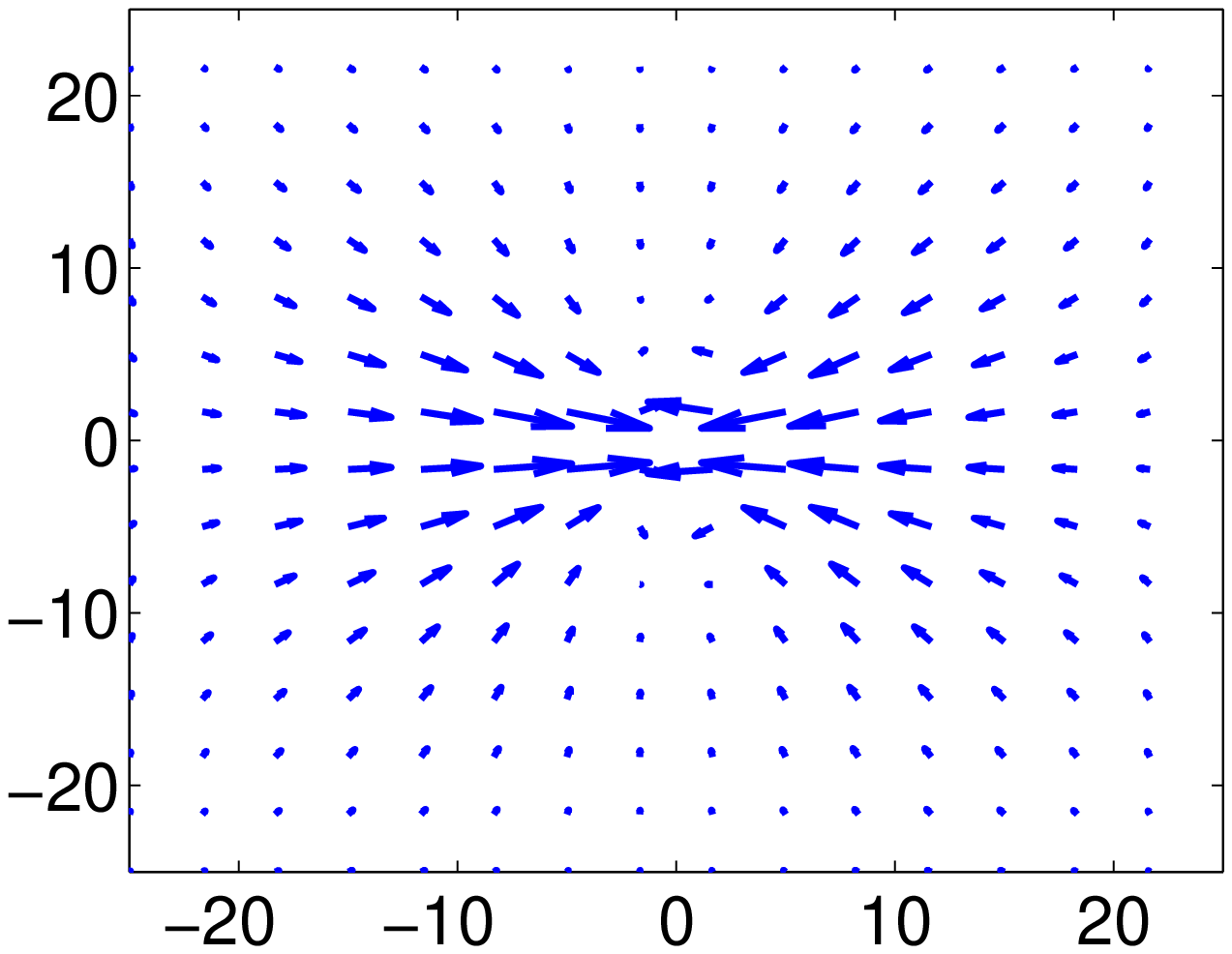}~~~&
\includegraphics[width=0.75\columnwidth]{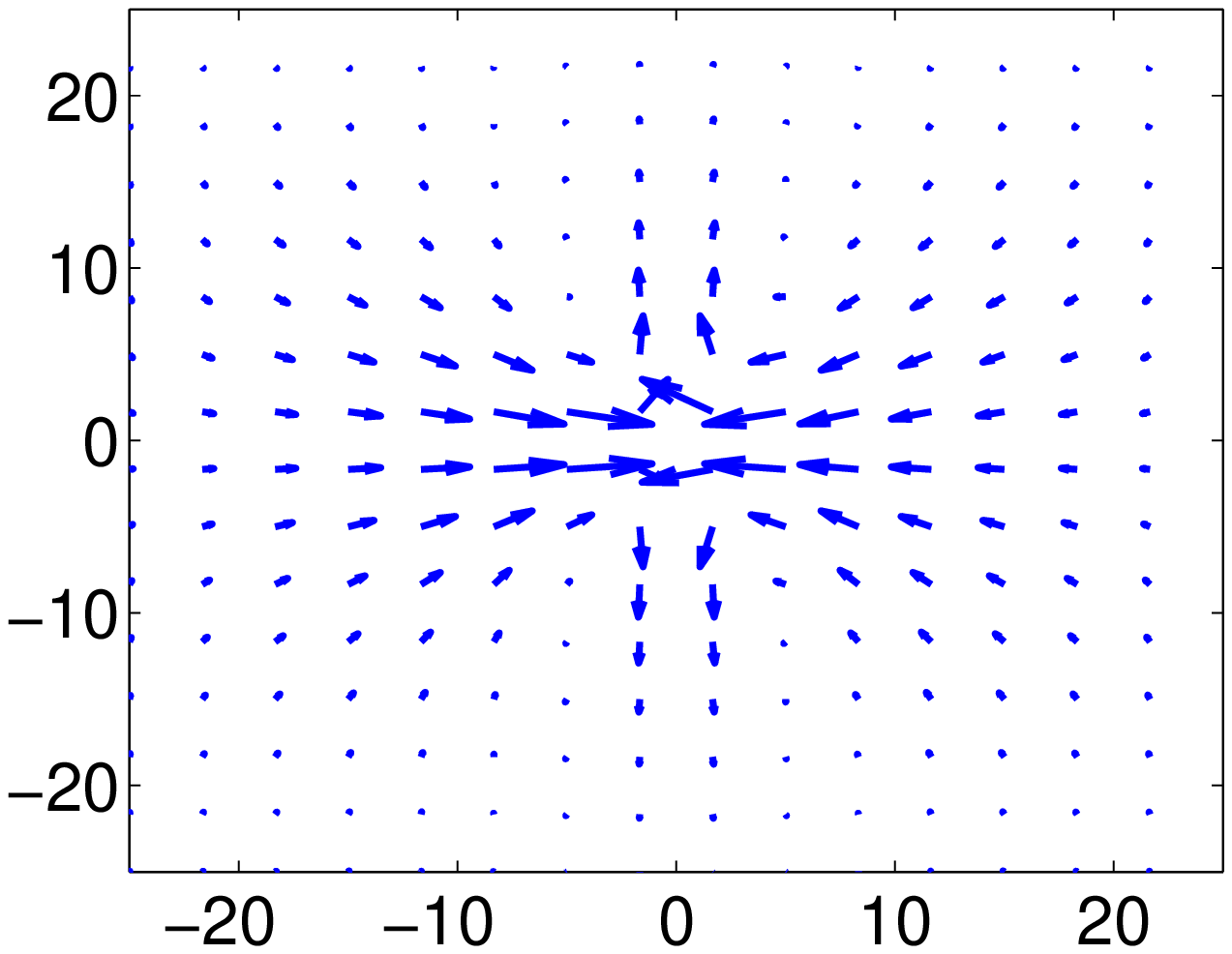}
\end{array}$
\caption{ The profile of a supercurrent  for the $s+is$ state. We set the parameter $\epsilon$, related to external
breaking of $C_4$ symmetry, to be $\e=0.1\sqrt{a_1a_2}$, and $0.3\sqrt{a_1a_2}$ in left and right panels, respectively. We also set $a_h= 0.5\sqrt{a_1a_2}$, $\delta=0.1/r_0$, and the sample size $L=25 r_0$ where, we remind, $r_0$ is related to impurity-induced correction to the prefactor $\alpha$ in the quadratic part in the free energy via $\alpha_{\text{imp}} (\br) \propto e^{-(r/r_0)^2}$. The asymmetry in the current pattern from $x$ to $y$ is due to the applied strain.}
\end{figure*}

{\it $s+is$ superconductor}~~
We now apply the same phenomenological description to $s+is$ superconductor with order parameters $\Delta_s$ and $\Delta_{s'}$. The generic form of $\mathcal{F}$
remains the same as in Eq. (\ref{eq:sidfree}), we only need to replace $\Delta_d$ by $\Delta_{s'}$, $\gamma_d$ by $\gamma_{s'}$, $\gamma_{sd}$ by $\gamma_{ss'}$, and  change the sign between $D_x$ and $D_y$ parts of the mixed term, which for $C_4$-preserving $s+is$ superconductor becomes
\beq
\gamma_{s s'} \left[(\vec{D}_x \Delta_s)^*\vec{D}_x \Delta_{s'} + (\vec{D}_y \Delta_s)^*\vec{D}_y\Delta_{s'}~+~\text{c.c.}\right],
\label{n_2}
\eeq
Such term is allowed by symmetry as it again depends only on the relative phase of $\Delta_s$ and $\Delta_{s'}$  and is invariant under gauge transformation and TR operation. However, in distinction to $s+id$ case, there is no possibility  to compensate the change of the phase of  $\Delta_s'$ by $\pi$ by a rotation of the coordinate frame by $90^o$.  As a result, the contributions from this term must vanish in the $s+is$ state, hence it should not give rise to a circulating current. To see this explicitly, we calculate the $\bj$ using the cross-term with the derivatives in the form of Eq. (\ref{n_2}). The current is still given by Eq. (\ref{eq:solutions1}), but now  $t_{\pm}=\alpha_{\text{imp}}/(k^2+ \delta^2)(1\pm\tilde{\gamma})$. Substituting this into (\ref{eq:solutions1}) we find that $\bj$ vanishes, as expected:
\beq\label{eq:currents3}
\bj=\frac{j_0}{2}\frac{\gamma_s-\gamma_{s'}}{\sqrt{\gamma_s\gamma_{s'}}}\vec\partial\frac{\alpha_{\text{imp}}}{k^2+ \delta^2}\left(\frac{-2\tilde\gamma}{1-\tilde\gamma^2} + \frac{2\tilde\gamma}{1-\tilde\gamma^2}\right)=0.
\eeq

We now  substantiate this reasoning with the analysis of the minimal
band-resolved model
which displays $s+is$ superconductivity. The model consists of two $\Gamma$-centered hole pockets and two electron pockets centered around $(0,\pi)$ and $(\pi,0)$ in the one-Fe Brillouin zone, as shown in Fig. \ref{fig:FS}.
(In the physical two-Fe Brillouin zone, these two electron pockets are centered at the same $(\pi,\pi)$ point and  hybridize into inner and outer pockets.)
%
We label the gap functions on hole pockets as $\Delta_{h_1}, \Delta_{h_2}$ and use $\Delta_{e_1}$ and $\Delta_{e_2}$ for inner and outer electron pockets.
As discussed in Ref. \onlinecite{maiti_13}, an $s+is$ can be realized at large hole doping where $\Delta_{h_1} = \Delta_1 e^{i \alpha_1}$, $\Delta_{h_2} = \Delta_2 e^{i \alpha_2}$, and $\alpha_1-\alpha_2$ is a fraction of $\pi$. The spatially varying part of the free energy for this state is
\bea\label{eq:Free energy} \mathcal{F}_s&=&  a_1 |\vec{D}
\Delta_{h_1}|^2 + a_2 |\vec{D}
\Delta_{h_2}|^2 + a_3 |\vec{D} \Delta_{e_1}|^2 + \nonumber\\
&&a_4 |\vec{D} \Delta_{e_2}|^2 +a_h\left[(\vec{D} \Delta_{h_1})^*\cdot(\vec{D} \Delta_{h_2})~+~c.c\right]\nonumber\\
&&+ a_{h_1e_1}\left[(\vec{D} \Delta_{e_1})^*\cdot(\vec{D}
\Delta_{h_1})~+~c.c\right]~+~\left[e_1 \rightarrow e_2\right]\nonumber\\
&&+ a_{h_2e_1}\left[(\vec{D} \Delta_{e_1})^*\cdot(\vec{D}
\Delta_{h_2})~+~c.c\right]~+~\left[e_1 \rightarrow e_2\right]\nonumber\\
&&+ a_{e}\left[(\vec{D} \Delta_{e_1})^*\cdot(\vec{D}
\Delta_{e_2})~+~c.c\right]\eea
The terms with products $(\vec{D} \Delta_{i})^*\cdot(\vec{D} \Delta_{j})$ with $i \neq j$  depend on the relative phases of $\Delta_{i}$ and $\Delta_{j}$ and  their structure reproduces that in Eq. (\ref{n_2}).

Differentiating the free energy with respect to vector potential we obtain the current $\vj=-\frac{\partial\mathcal{F}_s}{\partial \vec{A}} |_{\vec{A}=0}$  in the form
\bea\label{eq:currents}
i\frac{j_x}{2e/c}&=&a_1\Delta_{h_1}^*\check\partial_x\Delta_{h_1} +a_2\Delta_{h_2}^*\check\partial_x\Delta_{h_2}
+a_3\Delta_{e_1}^*\check\partial_x\Delta_{e_1} \nonumber \\
&&+a_4\Delta_{e_2}^*\check\partial_x\Delta_{e_2} +a_h\left[ \Delta_{h_1}^*\check\partial_x\Delta_{h_2}~-~\text{c.c.}\right]\nonumber\\
&&+a_{h_1e_1}\left[ \Delta_{e_1}^*\check\partial_x\Delta_{h_1}~-~\text{c.c.} \right]~+~\left[e_1 \rightarrow e_2\right]\nonumber\\
&&+a_{h_2e_1}\left[ \Delta_{e_1}^*\check\partial_x\Delta_{h_2}~-~\text{c.c.} \right]~+~\left[e_1 \rightarrow e_2\right]\nonumber\\
&&+a_e\left[
\Delta_{e_1}^*\check\partial_x\Delta_{e_2}~-~\text{c.c.}\right],\eea
where $g\check\partial h\equiv g\partial h-h\partial g$. The expression for $j_y$ is obtained by interchanging $\check\partial_x$ by $\check\partial_y$.

We perform the same computational steps as for $s+id$ superconductor: introduce an impurity at ${\br}=0$ and expand the order parameters to linear order around the homogeneous solution for $s+is$ state. We first consider the simplest case  when we treat the two hole pockets and the two electron pockets as identical  in the spatially varying part of the free energy,  i.e., set $a_1=a_2$, $a_3=a_4$, $a_{h_i e_j} = a_{he}$ in Eq. (\ref{eq:Free energy}).  The homogeneous solution for identical hole and identical electron  pockets is\cite{maiti_13} $\Delta_{h_1} = \Delta e^{i \alpha}$, $\Delta_{h_2} = \Delta e^{-i \alpha}$, $\Delta_e=-\eta\Delta$, where $\eta$ and $\alpha$ are functions of doping. For the critical doping where $T_c^{TRSB}$ is the largest, $\eta=\sqrt{2}$ and $\alpha=\pi/4$. We expand the gaps as
$\Delta_{h_1} = \Delta e^{i\alpha}(1+ m_1 + i\phi_1),~\Delta_{h_2} =\Delta e^{-i\alpha} (1+ m_2 + i\phi_2),~
\Delta_{e_1} = \Delta_{e2} = -\eta\Delta (1+ m_e + i\phi_e)$, substitute the expansion into (\ref{eq:currents}) and obtain for the current
\bea\label{eq:scurrent}
j_x &=& j_0\left[ (a_1 + a_h\cos2\alpha-2a_{he}r\cos\alpha )\partial_x(\phi_1+\phi_2)\right.\nonumber\\
&&+ (2a_3r^2+2a_er^2-4a_{he}r\cos\alpha) \partial_x\phi_e\nonumber\\
&&+\left. (a_h\sin2\alpha -2a_{he}r\sin\alpha)\partial_x(m_1-m_2)\right],
\eea
and $j_y$ is obtained by interchanging $x$ and $y$. Minimizing $ \mathcal{F}_s$ with respect to variations $m_i, \phi_i, m_e$ and $\phi_e$ we obtain that $\phi_e =0$ and $m_1=m_2$, $\phi_1 = -\phi_2$, where
\bea
m_1 + i\phi_1 &=& \int\frac{d^2k}{(2\pi)^2}e^{i\bk\cdot\br}\frac{a_3 +a_e + 2a_{he}}{G}e^{i\alpha}\frac{-\tilde\alpha_{\text{imp}}}{k^2+\delta^2}, \nonumber
\eea
and $G \equiv (a_1+ a_h)(a_3+a_e) -4a^2_{he}\eta$. Substituting this into Eq. (\ref{eq:scurrent}) we see that each term in (\ref{eq:scurrent}) vanishes, i.e., $\bj=0$. We extended the analysis to non-equivalent hole pockets and
to distinct inner and outer electron pockets, and obtained the same result~\cite{comm}: spontaneous supercurrent around an impurity vanishes. In these cases, individual contributions to the current do not vanish but the variations of the gap magnitudes ($\partial_x m$-terms) cancel the variations of the phases ($\partial_x{\phi}$ terms).

{\it $s+is$ superconductor under external strain}~~ The comparative analysis of $s+id$ and $s+is$ cases shows a way how one can  generate supercurrents in an $s+is$ superconductor --- one has to externally break C$_4$ rotational symmetry. This can be achieved, for example, by application of a small uniaxial strain.
For Eq. (\ref{n_2}) breaking of $C_4$  implies that $\gamma_{s,s'}$ along $x$ and $y$ directions become non-equivalent:
$\gamma^{x}_{ss} = \gamma_{ss} + \gamma^*_{ss}$ and  $\gamma^{y}_{ss} = \gamma_{ss} - \gamma^*_{ss}$
 Then  the cross-term becomes the sum of (\ref{n_2}) and
 the new term
\beq
 \gamma^*_{ss'} \left[(\vec{D}_x \Delta_{s})^*\vec{D}_x \Delta_{s'} - (\vec{D}_y \Delta_{s})^*\vec{D}_y\Delta_{s'}+\text{c.c.}\right],
\label{n_3}
\eeq
This last term has the same form as for $s+id$ superconductor and should give rise to a supercurrent.
Another way to state the same is to recall that breaking of C$_4$ mixes $s-$wave and $d-$wave channels, i.e. $s-$wave gaps change from $\Delta_{a}$ to $\Delta_{a,1} + \epsilon \Delta_{a,2} \cos 2\theta_a$, where $a = s, s'$. Selecting the products $\Delta_{s,1} \Delta_{s',2}$ and  $\Delta_{s',1} \Delta_{s,2}$,  we recover Eq. (\ref{n_3}) with $\gamma^*_{s s^{'}} \propto \epsilon$. Note in passing that in KFe$_2$As$_2$ $\epsilon$ and $\gamma^*_{s s'}$ are additionally enhanced because $s-$wave and $d_{x^2-y^2}$ channels are nearly degenerate~\cite{maiti_11,s+id}.

We now apply this reasoning to our model for $s+is$ superconductor. To simplify the presentation, we assume that the effect of strain is the strongest on the $a_h$ term in $\mathcal{F}_s$, set $a_{h}^{x}=a_h+\e$ and $a_{h}^{y}=a_h-\e$, and ignore $a_{h_i e_j}$ and $a_e$ terms which involve electron pockets. Within this approximation, the currents are given by
\bea\label{eq:cs}
&&j_x=j_0\left[ \partial_x (a_1\phi_1 + a_2\phi_2) + (a_h+\e)\cos2\alpha~\partial_x(\phi_1+\phi_2)\right.\nonumber\\
&&\left.+ (a_h+\e)\sin2\alpha~\partial_x(m_1-m_2) \right],
\eea
and
$j_y$
is obtained by interchanging $x\rightarrow y$ and $\e\rightarrow -\e$.

The coordinate-dependent functions $m_{j}$ and $\phi_j$ ($j=1,2$) are obtained by minimizing $\mathcal{F}_s$ with respect to fluctuations. Performing the same calculations as before we obtained
\bea\label{eq:form}
&&j_x=2j_0\e\sin\alpha\frac{a_1-a_2}{\sqrt{a_1a_2}}\partial_x\int\frac{d^2k}{(2\pi)^2}e^{i\bk\cdot\br}\left( \frac{\tilde\alpha_{\text{imp}}k_y^2}{Z}\right),\nonumber\\
&&j_y=-2j_0\e\sin\alpha\frac{a_1-a_2}{\sqrt{a_1a_2}}\partial_y\int\frac{d^2k}{(2\pi)^2}e^{i\bk\cdot\br}\left( \frac{\tilde\alpha_{\text{imp}}k_x^2}{Z}\right),\nonumber\\
\eea
where $Z=a_1a_2k^4 - [a_hk^2 +\e (k_x^2-k_y^2)]^2$. We see that the current is non-zero
only when $\alpha \neq 0, \pm \pi$
, $\e \neq 0$  (i.e. when  TRS {\it and} C$_4$ symmetry are simultaneously broken) and $a_1\neq a_2$ (i.e. when the two hole pockets have inequivalent fluctuations (this is analogous
to the condition  $\gamma_s \neq \gamma_{d}$ for
$s+id$ superconductor; see Eq. (\ref{eq:currents2})). We also emphasize that, although electron pockets do not contribute to ${\bj}$ in Eq. (\ref{eq:form}), they are essential to have  $\alpha\neq 0,\pm\pi$. Note also that the continuity equation $\vec\nabla\cdot{\bj} =0$ is satisfied.

We show the current profile for ${\bj}$ given by Eq. (\ref{eq:form}) in Fig \ref{fig:currb}.  It is quite similar to the one for $s+id$ superconductor. Superconductivity with $s+is$ symmetry has been proposed for KFe$_2$As$_2$ and we call for $\mu$SR measurements in this material under external strain.

{\it Conclusions} In this paper we analyzed $s+is$ state proposed as a candidate superconducting state for strongly hole-doped Ba$_{1-x}$K$_x$Fe$_2$As$_2$. This state breaks TRS but does not break any other discrete symmetry. We addressed the issue whether  TRS breaking alone can generate spontaneous supercurrents, which can potentially  be detected in $\mu$SR experiments.   Our conclusion is that there are no currents if only TRS is  broken. We argue, however, that spontaneous supercurrents do emerge in an $s+is$ superconductor  if the system is put under external strain and C$_4$ lattice rotation symmetry is externally broken.

We thank C. Kallin, W.-C. Lee, S. Raghu, J. Sonner, and R. Thomale for fruitful discussions. SM is a Dirac Post-Doctoral Fellow at the National High Magnetic Field Laboratory, which is supported by the NSF via Cooperative agreement No. DMR-1157490, the State of Florida, and the DOE. AVC is supported by the DOE grant DE-FG02-ER46900. The authors acknowledge the support of KITP (via Grant No. NSF PHY11-25915), where the bulk of this work has been performed.

\newpage
\begin{widetext}

\section*{Supplementary Material}

\subsection{Free energy analysis and spontaneous supercurrents in $s+id$ state}

We discuss here in more detail the properties of the Free energy $\mathcal{F}$ presented in Eq. (1) of the main text (MT) and derive the expression for the supercurrent -- Eq. (5) of the MT.  We first discuss the properties of the homogenous part of the Free energy given by:
\bea\label{eq:F_homo}
\mathcal{F}_{h}&=& \alpha_s |\Delta_s|^2 + \alpha_d |\Delta_d|^2 + \beta_1|\Delta_s|^4 + \beta_2|\Delta_d|^4 + \beta_3|\Delta_s|^2|\Delta_d|^2\nonumber\\
&&+\beta_4\left( \Delta_s^*\Delta_s^*\Delta_d\Delta_d~+~\text{c.c.} \right)
\eea
We set the coefficients of the quadratic term to be the same
 because we want to tune to the point where the time-reversal symmetry breaking (TRSB) state starts right at $T_c$. If $\alpha_d$ and $\alpha_s$
 differ a bit, TRSB state still emerges, but at $T < T_c$.

 Some properties of $\mathcal{F}$ at $\alpha_d = \alpha_s = \alpha$ are\\
(1) TRSB is realized only when $\beta_4>0$.\\
(2) $|\Delta_s|^2/|\Delta_d|^2~ = ~ (2\beta_2+2\beta_4-\beta_3)/(2\beta_1+2\beta_4-\beta_3)$.\\
(3) The phases of the order parameters differ by $\pm \pi/2$. We will set the cumulative phase such that $\phi_s=0$ and $\phi_d=\pm \pi/2$.
For the sake of convenience we choose $\beta_1=\beta_2$ in which case $|\Delta_s|=|\Delta_d|$.

If there is an impurity at the origin, there will be a spatially varying part of the Free energy given by:
\bea\label{eq:Free energy_space} \mathcal{F}_s&=&  \gamma_s |\vec{D}
\Delta_s|^2 + \gamma_d|\vec{D}
\Delta_d|^2 +\gamma_{sd} \left((\vec{D}_x \Delta_s)^*\vec{D}_x \Delta_d- (\vec{D}_y \Delta_s)^*\vec{D}_y\Delta_d~+~\text{c.c.}\right),\nonumber\\\eea
where $\vec{D}\equiv -i\vec{\partial} -\frac{2e}{c}\vec{A}$.
We follow Ref. [\onlinecite{wu}] and model the effect of impurity  by replacing $\alpha_2$ by $\alpha_2+\alpha_{\text{imp}}(\br)$, where $\alpha_{\text{imp}}=\alpha_0 e^{-(r/r_0)^2}$.

Minimizing $\mathcal{F}$ leads to the coupled Ginzburg Landau(GL) equations-
\bea\label{eq:coupled GL}
\alpha_2\Delta_s + 2\beta_1|\Delta_s|^2\Delta_s + \beta_3|\Delta_d|^2\Delta_s + \beta_4\Delta_d^2\Delta_s^* +
\gamma_s (D_x^2+D_y^2) \Delta_s+ \gamma_{sd}(D_x^2-D_y^2)\Delta_d&=&0\\
(s\leftrightarrow d).~~~~~~~~~~~~~~~~~~~~~~~~~~~~~&&
\eea
The current is
\bea\label{eq:current_def}
\bj&=&\bj_s + \bj_d + \bj_{sd}\\
\bj_s&=&-i\frac{2e}{c}\gamma_s\left[\Delta_s^*\vec\partial\Delta_s\right]~+~\text{c.c.}\\
\bj_d&=&-i\frac{2e}{c}\gamma_d\left[\Delta_d^*\vec\partial\Delta_d\right]~+~\text{c.c.}\\
\bj_{sd}&=&-i\frac{2e}{c}\gamma_{sd}\left[(\Delta_s^*\partial_x\Delta_d+\Delta_d^*\partial_x\Delta_s)\hat{x}-(x\rightarrow y)\right]~+~\text{c.c.}
\eea
The current $\bj$ needs to satisfy the continuity equation $\vec{\nabla}\cdot\bj=0$.

This non-linear system of partial differential equations can be solved numerically. However,to gain insight, we may choose to expand in the spatial variations induced by the impurity around the homogenous TRSB state and analyze a linearized version of the GL equations. As we shall see, this gives us an excellent intuitive picture of how the solutions for $\Delta_{s,d} (r)$ look like.  We expand $\Delta_s = |\Delta_s| e^{i\phi_s}$ and $\Delta_d = |\Delta_d|e^{i\phi_d}$ near homogeneous TRSB state $\Delta_s =|\Delta|, \Delta_d = i \Delta$ as
\bea\label{eq:subs}
\Delta_s&=&\Delta + m_s + i\phi_s\Delta, ~~\text{and}\nonumber\\
\Delta_d&=&i(\Delta + m_d + i\phi_d\Delta).
\eea
where the variations of the gap magnitudes($m_s$ and $m_d$) and the phases($\phi_{s}$ and $\phi_d$) are the spatially varying parts  due to the presence of an impurity. Let us now rescale variables for convenience:

\bea\label{eq:rescaling}
&&\alpha\Delta\rightarrow \alpha;~\beta\Delta^3\rightarrow \beta;~\frac{m_s}{\Delta}\rightarrow m_s;~\frac{m_d}{\Delta}\rightarrow m_d;~\frac{r}{r_0}\rightarrow r;~\gamma\Delta r_0^{-2}\rightarrow \gamma;~
\frac{4e\Delta r_0}{c}\rightarrow j_0.
\eea
If the impurity perturbation is weak, we can also neglect the feedback effect of $\vec A$ and set it to zero in the calculations of $m_i$, $\phi_i$ ($i = s,d$). We will see that this does not lead to any inconsistency. Substituting Eq. (\ref{eq:subs}) into the GL equations and expanding to linear order in $m_i$ and $\phi_i$ we obtain
\bea\label{eq:GL}
\left(
\begin{array}{cccc}
A-\gamma_s\partial^2&B&0&\gamma_{sd} \partial_d^2\\
B&A-\gamma_d\partial^2&-\gamma_{sd} \partial_d^2&0\\
0&-\gamma_{sd} \partial_d^2&\tilde{A} - \gamma_s\partial^2&-\tilde{B}\\
-\gamma_{sd} \partial_d^2&0&\tilde{B}&-\tilde{A} + \gamma_d \partial^2
\end{array}
\right)\left(
\begin{array}{c}
m_s\\
m_d\\
\phi_s\\
\phi_d
\end{array}
\right)
&=&
\left(
\begin{array}{c}
-\alpha_{\text{imp}}\\
-\alpha_{\text{imp}}\\
0\\
0
\end{array}
\right).
\eea
where $A=\alpha_2 + 6\beta_1+ \beta_3 -\beta_4$; $B=2\beta_3 - 4\beta_4$; $\tilde{A}=\alpha_2 + 2\beta_1+ \beta_3 +\beta_4$; $\tilde{B}= 2\beta_4$; $\partial_d^2 = \partial_x^2+\partial_y^2$; and $\partial_d^2=\partial_x^2-\partial_y^2$. In the same approximation, the current is given by
\bea\label{eq:newcurrents}
\bj&=&-j_0\left[\gamma_s\vec{\partial}\phi_s + \gamma_d\vec{\partial}\phi_d + \gamma_{sd} \vec{\partial}_d(m_d-m_s)\right],
\eea
where $\vec{\partial}_d=\partial_x \hat{x} - \partial_y \hat{y}$. To move on with analytic calculations we assume that $r_0\ll l_0$, where $l_0=\text{min}\left\{\sqrt{\gamma/\alpha},\sqrt{\gamma/\beta\Delta^2}\right\}$. This inequality guarantees that  $\gamma\gg A,B,\tilde{A},\tilde{B}$.
Then the system of GL equations decouples into two $2\times2$ sets. One is
\bea\label{eq:GL2}
\left(
\begin{array}{cc}
-\gamma_d\partial^2&-\gamma_{sd} \partial_d^2\\
-\gamma_{sd} \partial_d^2&-\gamma_s\partial^2
\end{array}
\right)\left(
\begin{array}{c}
m_d\\
\phi_s
\end{array}
\right)
&=&
\left(
\begin{array}{c}
-\alpha_{\text{imp}}\\
0
\end{array}
\right),
\eea
and in the other set we replace $d\leftrightarrow s)$; and $\phi_d$ by $-\phi_s$. This system of partial differential equations has a particular solution induced by $\alpha_{\text{imp}}$  and a general solution for $\alpha_{\text{imp}} =0$. We can
set the general solution to zero by arguing that in the absence of the impurity, the system has no spontaneous currents and since the impurity potential is short ranged,
the particular solution, which gives the response to $\alpha_{\text{imp}}$, can then be found using the Fourier Transform and matrix inversion. Performing the calculation we obtain (in the real space):
\bea\label{eq:solutions}
\gamma_d m_d = \gamma_s m_s &=& -\frac12\left(t_+ + t_-\right), \nonumber\\
\sqrt{\gamma_s\gamma_d} \phi_d =-\sqrt{\gamma_s\gamma_d} \phi_s&=& \frac12 \left( t_+ - t_- \right),
\eea
where
\beq\label{eq:def}
t_{\pm}=\int\frac{d^2k}{(2\pi)^2}e^{i\bk\cdot\br}\frac{\tilde\alpha_{\text{imp}}}{k_x^2(1\pm\tilde\gamma) + k_y^2(1\mp\tilde\gamma) + \delta^2},
\eeq
where $\tilde\gamma \equiv \gamma_{sd}/\sqrt{\gamma_s\gamma_d}$, and $\tilde\alpha_{\text{imp}}$ is the fourier transform of $\alpha_{\text{imp}}$. The parameter $\delta$ equals to zero in this calculation, but this is an artifact because
we have ignored $A,B...$ relative to $\gamma$'s. If we keep $A, B..$, we find that $\delta\neq 0$. We didn't compute $\delta$ explicitly and will use it just as a parameter which cuts off infrared singularity.

Substituting $m_d, m_s, \phi_d$ and $\phi_s = - \phi_d$ into the expression for the current we obtain
\beq\label{eq:currents2}
\bj=\frac{j_0}{2}\frac{\gamma_s-\gamma_d}{\sqrt{\gamma_s\gamma_d}}\left[\vec{\partial}(t_+-t_-) + \tilde\gamma\vec{\partial}_d(t_++t_-)\right].
\eeq
This is the result which we  presented in the MT and plotted in Fig. 1 there.  To check if this current satisfies the continuity equation, let us go back to momentum space and look at $\bk\cdot\bj$.  We immediately find that $\bk\cdot\bj=-(k_x^2+k_y^2)(t_+-t_-)-\tilde\gamma(k_x^2-k_y^2)(t_+-t_-)=0$.
Like we said in the MT, our result is quite consistent with the numerical results for $\bj$ obtained in Ref. [\onlinecite{wu}].
The message then is that the  spontaneous current can be fully understood by studying the linearized GL equations and looking only at the spatially varying part of $\mathcal{F}$. We also emphasize that the existence of a non-zero $\bj$ is entirely due to the presence of $\gamma_{sd}$ term in the Free energy. Indeed,
if $\gamma_{sd}=0$, then  ${\tilde \gamma} =0$ and
$t_+ = t_-$, hence $\bj =0$.

We remind that, in this analysis, there is a characteristic length scale $r_0$, that we expressed our lengths in. We repeat the assumptions  that we made: \\
(1)$r_0\gg k_F^{-1} \rightarrow$. This allows us to consider slow variations of our variables as functions of $\vec{r}$.\\
(2) $r_0\ll l_0$. This justifies dropping $A, B, {\tilde A}, {\tilde B}$ compared to $\gamma$, i.e., we can neglect the homogeneous part of the Free energy in solving GL equations. \\
Combining the two approximations, we find that our analytic consideration is valid when  $k_F^{-1}\ll r_0\ll l_0$.

\subsection{Test cases for the lack of supercurrents in $s+is$ state without external strain}

Here we consider in detail the microscopic model discussed in the MT to show that there are no supercurrents induced in $s+is$ state. It was shown in the MT, that for the case when the spatially varying (fluctuating) part of the Free energy does not distinguish between the two hole pockets and between the two electron pockets, there are no supercurrents. Here we consider three more generic cases: (a) equivalent fluctuations of superconducting gaps on hole pockets and inequivalent gap fluctuations on the electron pockets; (b) inequivalent gap fluctuations on the hole pockets and equivalent gap fluctuations on the electron pockets; and (c) inequivalent gap fluctuations on the two hole pockets {\it and} on the two  electron pockets.

We emphasize that in all these cases we expand around the homogeneous TRSB state with
$\Delta_{h_1} =\Delta e^{i\alpha}, \Delta_{h_2} =\Delta e^{-i\alpha}$ and  $\Delta_{e_1} = \Delta_{e_2} = -\eta\Delta$.
The purpose to perform these checks is to show that, unless a spatial symmetry (in this case C$_4$) is broken, there cannot a supercurrent even if the various bands interact differently with the impurities. The general expansion around the TRSB state is:
\bea\label{eq:sexpand}
\Delta_{h_1}&=&\Delta e^{i\alpha}(1+ m_1 + i\phi_1)\nonumber\\
\Delta_{h_2}&=&\Delta e^{-i\alpha} (1+ m_2 + i\phi_2)\nonumber\\
\Delta_{e_1}&=&-\eta\Delta (1+ m_{e_1} + i\phi_{e_1})\nonumber\\
\Delta_{e_2}&=&-\eta\Delta (1+ m_{e_2} + i\phi_{e_2}).
\eea

\emph{Case(a): equivalent gap fluctuations on hole pockets and inequivalent gap fluctuations on the electron pockets:}
To realize this case, we take $a_{h_ie_1}\neq a_{h_ie_2}$, but $a_{h_1e_i}= a_{h_2e_i}$.
This case is realized when we include into consideration the  hybridization between the two elliptical electron pockets, which gives rise to the splitting of the two electron pockets in the folded Brillioun zone into non-equivalent inner and outer pockets, each of which respects C$_4$ symmetry. Note from Eq. (11) in the MT that $a_{1,2,3,4}$ do not really affect qualitatively our results as they always enter in combination with $a_h$ or $a_e$; thus we may even set them to zero without qualitatively affecting the result. With this assumption, the current is:
\bea\label{eq:scurr}
j_x&=&j_0\left[ (a_h-(a_{he_1}+a_{he_2})\eta\cos\alpha)\partial_x(\phi_1+\phi_2)+(a_h\sin2\alpha + (a_{he_1}+a_{he_2})\eta\sin\alpha)\partial_x(m_1-m_2)\right.\nonumber\\
&&\left.+(a_e\eta^2-2a_{he_1}\eta\cos\alpha )\partial_x\phi_{e_1} +  (a_e\eta^2-2a_{he_2}\eta\cos\alpha )\partial_x\phi_{e_2}) \right],
\eea
and $j_y=(x\rightarrow y)$. The GL equations are:
\bea\label{eq:GL2_1}
\left(
\begin{array}{cccc}
0&a_h&-a_{h_1e_1}&-a_{h_1e_2}\\
a_h&0&-a_{h_2e_1}&-a_{h_2e_2}\\
-a_{h_1e_1}&-a_{h_2e_1}&0&a_e\eta\\
-a_{h_1e_2}&-a_{h_2e_2}&a_e\eta&0
\end{array}
\right)\left(
\begin{array}{c}
g_1\\
g_2\\
g_3\\
g_4
\end{array}
\right)
&=&-R
\left(
\begin{array}{c}
1\\
1\\
1\\
1
\end{array}
\right),
\eea
where $g_1=e^{i\alpha}(m_1 + i\phi_1)$, $g_2=e^{-i\alpha}(m_2 + i\phi_2)$, $g_3=\eta(m_{e_1} + i\phi_{e_1})$, $g_4=\eta(m_{e_2} + i\phi_{e_2})$; and
\beq\label{eq:ss}
R\equiv\int\frac{d^2k}{(2\pi)^2}e^{i\bk\cdot\br}\frac{\tilde\alpha_{\text{imp}}}{k^2+\delta^2}.
\eeq
After matrix inversion, we get $\phi_{e_1,e_2}=0$ (by looking at the imaginary part of the solutions for $g_i$). Since $a_{h_1e_i}=a_{h_2e_i}$ we find $m_1=m_2$ and $\phi_1=-\phi_2$. This leads to the vanishing of the supercurrent.

\emph{Case(b) inequivalent gap fluctuations of the hole pockets and equivalent gap fluctuations on the electron pockets}: In this case, we treat electron pockets as circular and neglect the hybridization between them, but no longer assume that fluctuations on the two hole pockets  are identical. Since we are probing here the effect of inequivalent fluctuations of the hole pockets, the term $a_h$, which was already included in previous analysis, may be safely set to zero. The same argument allows us to set $a_e=0$.  In distinction with the previous case, we, however, need to
keep $a_1$ and $a_2$. We found in the MT that $a_1+ a_2$ does not  contribute to the current. We use this and set from the beginning $a_1=\tilde\e$ and $a_2=-\tilde\e$. We keep $a_{h_ie_1}=a_{h_ie_2}$, but set $a_{h_1e_i}\neq a_{h_2e_i}$. The supercurrent  is given by (using the definition given in the MT)
\bea\label{eq:currr}
j_x&=&j_0\left[ \tilde\e~\partial_x(\phi_1-\phi_2) + 2a_{h_1e}\left\{-\cos\alpha~\partial_x(\phi_1 + \frac{\tilde\phi_{e}}{2})-\sin\alpha~\partial_x(m_1-\frac{\tilde{m}_{e}}{2}) \right\}  \right.\nonumber\\
&&\left.+2a_{h_2e}\left\{-\cos\alpha~\partial_x(\phi_2 + \frac{\tilde\phi_e}{2})+\sin\alpha~\partial_x(m_2-\frac{\tilde{s}_e}{2}) \right\}     \right], \nonumber\\
j_y&=&(x\rightarrow y),
\eea
where $\tilde{\phi}_e=\phi_{e_1} + \phi_{e_2}$ and $\tilde{m_e}=m_{e_1} + m_{e_2}$. The GL equations now read:
\bea\label{eq:GL5}
\left(
\begin{array}{ccc}
\tilde\e&0&-a_{h_1e}\\
0&-\tilde\e&-a_{h_2e}\\
-a_{h_1e}-a_{h_2e}&-a_{h_1e}-a_{h_2e}&0\\
\end{array}
\right)\left(
\begin{array}{c}
g_1\\
g_2\\
\tilde{g}_3\\
\end{array}
\right)
&=&-R
\left(
\begin{array}{c}
1\\
1\\
1
\end{array}
\right),
\eea
where $\tilde{g}_e=g_{e_1} + g_{e_2}$.  The solution to the GL equations are:
\bea\label{eq:sol22}
m_1+i\phi_1&=&-R\frac{a^2_{h_1e}-a^2_{h_2e}-a_{h_1e}\tilde\e}{\tilde\e(a^2_{h_1e}-a^2_{h_2e})}e^{-i\alpha},\\
m_2+i\phi_2&=&R\frac{a^2_{h_1e}-a^2_{h_2e}-a_{h_2e}\tilde\e}{\tilde\e(a^2_{h_1e}-a^2_{h_2e})}e^{i\alpha},\\
\tilde{g}_e&=&R\frac{\tilde\e}{a^2_{h_1e}-a^2_{h_2e}},\\
\tilde\phi_e&=&0.
\eea
Plugging this back into Eq. (\ref{eq:currr}) we see that the $\sin\alpha$ and $\cos\alpha$ terms of $m_{1,2}$ and $\phi_{1,2}$ cancel each other out, and the variations of fluctuations of the gap amplitude $\tilde{m}_e$ cancels the variation of the relative phase $\phi_1-\phi_2$. Thus there is no supercurrent.

\emph{Case(c) inequivalent gap fluctuations on the two hole pockets and on the two  electron pockets}: This case is a combination of cases (a) and (b) and it is realized if we set $a_{h_ie_1}\neq a_{h_ie_2}$. The current is then given by:
\bea\label{eq:currr2}
j_x&=&j_0\left[ \tilde\e~\partial_x(\phi_1-\phi_2) -\cos\alpha\left\{ (a_{h_1e_1}+a_{h_1e_2})\partial_x\phi_1 + (a_{h_2e_1}+a_{h_2e_2})\partial_x\phi_2)\right\} \right.\nonumber\\
&& -\sin\alpha\left\{ (a_{h_1e_1}+a_{h_1e_2})\partial_xm_1 - (a_{h_2e_1}+a_{h_2e_2})\partial_xm_2)\right\} \nonumber\\
&&\left. +\sin\alpha\left\{ (a_{h_1e_1}-a_{h_2e_1})\partial_xm_{e_1} + (a_{h_1e_2}-a_{h_2e_2})\partial_xm_{e_2})\right\}  \right] \nonumber\\
j_y&=&(x\rightarrow y).
\eea
The GL equations are given by:
\bea\label{eq:GL6}
\left(
\begin{array}{cccc}
\tilde\e&0&-a_{h_1e_1}&-a_{h_1e_2}\\
0&-\tilde\e&-a_{h_2e_1}&-a_{h_2e_2}\\
-a_{h_1e_1}&-a_{h_2e_1}&0&0\\
-a_{h_1e_2}&-a_{h_2e_2}&0&0
\end{array}
\right)\left(
\begin{array}{c}
g_1\\
g_2\\
g_3\\
g_4
\end{array}
\right)
&=&-R
\left(
\begin{array}{c}
1\\
1\\
1\\
1
\end{array}
\right),
\eea
The solutions are given by:
\bea\label{eq:sol22_1}
m_1+i\phi_1&=&-R\frac{a_{h_2e_1}-a_{h_2e_2}}{a_{h_1e_1}a_{h_2e_2}-a_{h_2e_1}a_{h_1e_2}}e^{-i\alpha},\\
m_2+i\phi_2&=& R\frac{a_{h_1e_1}-a_{h_1e_2}}{a_{h_1e_1}a_{h_2e_2}-a_{h_2e_1}a_{h_1e_2}}e^{ i\alpha},\\
\eta m_{e_1}&=&-R\frac{  a_{h_1e_2}a_{h_2e_2}(a_{h_2e_1}+a_{h_1e_1}) -a^2_{h_1e_2}a_{h_2e_1}-a^2_{h_2e_2}a_{h_1e_1} +(-a_{h_1e_1}a_{h_1e_2}+ a^2_{h_1e_2}-a^2_{h_2e_2}+a_{h_2e_1}a_{h_2e_2})\tilde\e}{(a_{h_1e_1}a_{h_2e_2}-a_{h_2e_1}a_{h_1e_2})^2},\nonumber\\
&&\\
\eta m_{e_2}&=&-R\frac{a_{h_1e_1}a_{h_2e_1}(a_{h_2e_2}+a_{h_1e_2})-a^2_{h_1e_1}a_{h_2e_2}-a^2_{h_2e_1}a_{h_1e_2}+(-a_{h_1e_1}a_{h_1e_2}+ a^2_{h_1e_1}-a^2_{h_2e_1}+a_{h_2e_1}a_{h_2e_2})\tilde\e}{(a_{h_1e_1}a_{h_2e_2}-a_{h_2e_1}a_{h_1e_2})^2},\nonumber\\
&&\\
\phi_{e_1}&=&0,\\
\phi_{e_2}&=&0.
\eea
Plugging this back into Eq. (\ref{eq:currr2}), we once again find that the $\sin\alpha$ and $\cos\alpha$ terms of $m_{1,2}$ and $\phi_{1,2}$ cancel each other out, and the variations of the gap magnitudes  $m_3$ and $m_4$ cancel the variation of the relative phase $\phi_1-\phi_2$, leading to vanishing of the supercurrent.

Thus we have seen that neither the in-equivalence of gap fluctuations on the hole pockets nor that on the electron pockets
induces supercurrents in the $s+is$ state as long as C$_4$ symmetry is maintained.

\end{widetext}

\end{document}